  \documentclass[10pt]{article}
 \usepackage{epsfig}
\usepackage{latexsym,amssymb}
\usepackage{amsmath}

\begin{document}
\title{The effect of disorder on the fracture nucleation  process}
  \author{S. Ciliberto, A. Guarino and  R. Scorretti $^\clubsuit$
\\   Ecole Normale
Sup\'erieure de Lyon, Laboratoire de Physique ,\\
 C.N.R.S. UMR5672,  \\ 46, All\'ee d'Italie, 69364 Lyon Cedex
07,  France\\ }
\maketitle

\begin{abstract}
The statistical properties of failure are studied in a fiber
bundle model with thermal noise. We show  that  the macroscopic
failure is produced by a thermal activation of microcracks. Most
importantly the effective temperature of the system is amplified
by the spatial disorder (heterogeneity) of the fiber bundle. The
case of a time dependent force and the validity of the Kaiser
effects are also discussed. These results can give more insight to
the  recent experimental observations on thermally activated crack
and can be useful to study the failure of electrical networks.
\end{abstract}


\section{Introduction}

  Material failure is a widely
studied phenomenon  not only    for the very important
technological   applications but also for the fundamental
statistical aspects, which are not yet very well understood. Many
models  have been   proposed to give more insight into the
statistical analysis of   failure. Among the most studied ones we
can mention the fuse and   the non-linear spring networks
\cite{hermann,hemmer,andersen,vanneste,Coleman,Phoenix,zapperi},
which can reproduce several features of crack precursors
experimentally
  observed in heterogeneous materials   subjected to a
quasi-statically increasing  stress
\cite{lockner,anifrani,noiprl,noieps,petri,maes}.    Specifically
the power law behaviour  of the   acoustic emission observed in
several experiments close to the   failure point.     However the
non-linear spring  networks and  the other related    models, in
their standard formulation,      are unable to describe the
behaviour of a material   subjected to a creep-test, which
consists in submitting
   a sample to a constant stress till the failure time.
Creep-tests are widely used by engineers in order to estimate
     the sample lifetime as a function of the applied stress.
      Modified fuse networks  have been   proposed to explain
the finite lifetime of a sample   subjected to a constant stress.
In ref.\cite{vanneste} for example, the Joule effect has been
considered to explain delayed crack. Instead  in
ref.\cite{Coleman, Phoenix} phenomenological force dependence of
the fiber lifetime  has been proposed, but not very well justified
from a physical point of view.  These models do not explain the
recent experimental results on 2-D crystals\cite{pauchard},
gels\cite{bonn} and heterogeneous materials\cite{noi}.

These experiments show that the lifetime $\tau$ of a sample,
subjected to an imposed stress $P$ is well predicted by the
equation
\begin{equation}   \tau =\tau _o \exp{\left( \alpha \frac{\Gamma^d
\ Y^{(d-1)}}{K \ T_{eff} \ P^{(2d-2)}}\right) }
   \label{const}   \end{equation}     where $\tau_o$
    is a constant, $\Gamma $ the surface energy, $Y$   the Young
modulus, $K$ the Boltzmann constant $\alpha $ a constant,   which
depend on the geometry, $T_{eff}$ is an effective   temperature
and $d$ the dimensionality of the system.   Eq.\ref{const} was
first derived by Pomeau and Golubovic   \cite{pomeau,golubovic}
for $d=2$ and  $d=3$. It has been   generalized  to any $d$ in
ref. \cite {bonn}.  The main physical   hypothesis behind
eq.\ref{const} is that the macroscopic failure  of a material is
produced by a thermal activation of micro
cracks\cite{pomeau,golubovic}. In the original Pomeau's theory
  \cite{pomeau} $T_{eff}$ of eq.\ref{const} coincides with the
  thermodynamic temperature $T$ while,    experimentally,
$T_{eff}>> T$ \cite{pauchard,bonn,noi}.
  Eq.(\ref{const}) is
based on the idea that the sample failure is due to the   thermal
nucleation of one defect (microcrack) initially present inside
      the material\cite{pomeau}.   However, experimentally
it has been observed  that the macroscopic failure is   in some
way related to the nucleation of many defects; this important
point will be discussed at the end of this paper.   Using
reasonable values for the material constants in eq.\ref{const}
\cite{noi} one sees that thermal fluctuations are too small to
activate the nucleation of microcraks in times $\tau $ measured in
the experiments.   It has been measured that the temperature
needed to have the measured lifetimes $\tau $ should be of order
of several thousands Kelvin. Specifically it has been estimated
that  for wood $T_{eff} \ge 3000K$ \cite{noi} , in 2D crystals
$1000K<T_{eff}<2500K$ \cite{pauchard} and $T_{eff}>10^{10}K$ for
gels \cite{bonn}.  In several materials the lifetime $\tau $ of
the sample is not influenced, in the limit of experimental errors
(20 \%), by a  variation of the temperature $T$ from  $20$ to $90$
${{}^{\circ }}C$. In contrast, experiments on 2D-crystals
\cite{pauchard} show that $\tau $ depends on $T$. Furthermore the
lifetime statistics is close to Gaussian in
ref.\cite{pauchard,noi}. Thus  eq.\ref{const} seems to give the
right dependence on $P$ of  the sample lifetime but there are a
lot of experimental observations that are not completely in
agreement with a standard nucleation problem. To explain these
results it has been supposed \cite{noi} that the lifetime $\tau $
of the sample depends on the heterogeneity of the material and
that disorder in some way enhances \footnote{The term {\it
"enhances"} has to be understood in the sense that the effect of
the disorder is such that the system behaves like if the real
temperature is bigger} thermal fluctuations so that the nucleation
time of defects becomes of the order of the measured ones.

Motivated by these experimental observations and to check the
validity of our hypothesis we proposed
 \cite{noi-condmat,tesi-Guarino}  a model based on a  democratic
  fiber bundle model (DFBM) with thermal noise.
 The
DFBM  is the simplest spring network proposed  a long time ago by
Pierce \cite{pierce} to study cable failure.  The DFBM has been
studied  by several authors  \cite{hermann, hemmer, andersen,
hansen, kloster}, and it turns out  to be a quite realistic model
for long flexible cables, low-twist yarn and more generally for
composite heterogeneous materials. In a classical DFBM, which is a
deterministic model, it is not possible to observe stochastic
phenomena such as nucleation. Introducing thermal noise into the
system we transformed the DFBM in a stochastic model, so that
processes such as nucleation become observable.

  The purpose of
this paper is to extend the results of our previous work
\cite{noi-condmat,tesi-Guarino} and those obtained analytically by
Roux \cite{roux}. We will show  that, adding to the spring network
a noise, which plays the role of a temperature, it is possible to
reproduce the behavior of a material subjected to a creep-test.
Furthermore   we can prove that, for such a process, the noise is
"amplified"   by the network disorder.

  The paper is organized  as follows. In section 2 the modified DFBM
model is   introduced. In section 3 the statistical features of
the model are studied analytically. Specifically we compute how
failure time $\tau $ is related to the disorder of the material,
when  the bundle is subjected to a constant  force. In section 4
the results of section 3 are verified numerically. In section 5 we
study the lifetime in the general case of a bundle subjected to a
time dependent force $F(t)$. We also propose a formula, which
allows us to
 estimate the failure time $\tau $ by knowing the whole history
of the medium, that is to say,  the time dependence  of the
imposed force $F(t)$. In section 6 we give a comprehensive
explication of some facts related to Kaiser effect, with a
comparison between the experimental data and the numerical results
obtained with our model. Discussion and conclusions follow in
section 7.

\section{The model}

\begin{figure}
\centerline{\epsfysize=0.9\linewidth \epsffile{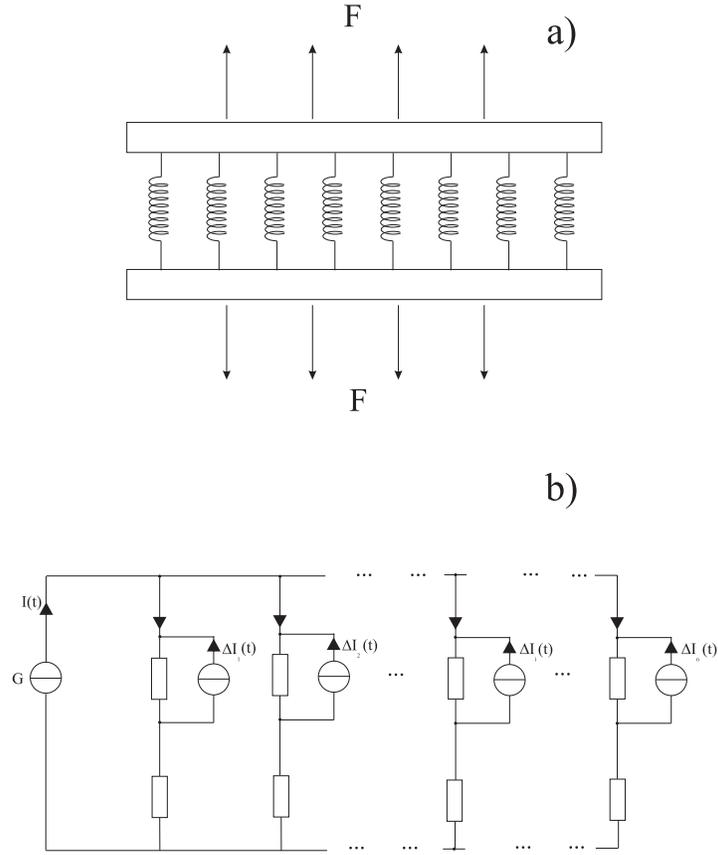}}
    \caption { a) Modified DFBM: $N$ fiber in parallel, with the edges
  fixed on a rigid support are subjected to an externally imposed
  force $F$ which is distributed democratically on the net, that is
  all fibers (not broken) are affected in the same way. Each fiber
  is also subjected to a random (zero mean, normally distributed),
  addictive force $\Delta f_i(t)$, where is intended that $\Delta
  f_i(t)$ is a realization of a white, time independent stochastic
  process. Name $n(t)$ the number of broken fibers at time $t$; we
  derive the following expression of local force $f_i$ for the i-th fiber: $%
  f_i=\frac F{N-n(t)}+\Delta f_i$. b) The equivalent of the fiber bundle model
  is the fuse network. One can think to our model as a fuse network where the Nyquist
  noise  of several resistances (current generators $\delta I_i$)
  are  the noise generators for each bond}
   \label{modello}
  \end{figure}

We simulate the behavior of an elastic heterogeneous medium
subjected to an imposed tensile force $F$ using a DFBM with
thermal noise. For this purpose,  we model the medium as a system
of $N$ parallel elastic fibers whose extremities   are fixed on a
rigid support as shown in fig. \ref{modello}.
   This model, widely studied in
the quasi static regime \cite{hermann,hemmer,andersen,vanneste},
is equivalent to $N$   springs in parallel subjected to a total
tensile force $F$.   Specifically we have studied the model by
using the   following rules:

\begin{itemize}

\item[{\sc  I}]  The external applied force $F$ produces a {\it local
force} $f_i$   on each fiber.   $F$ is democratically distributed
in the net:   $F=\sum_{i=1}^Nf_i$.

\item[{\sc II}]  The local force $f_i$ on the i-th fiber produces a {\it
local   strain} $e_i$. Being the media elastic, force and
displeacement are linked by the  Hooke's law:   \begin{equation}
f_i=Y\cdot e_i
  \end{equation}   where $Y$ is the stifness, which
is assumed to be the same   for all the fibers: $Y_i=Y$.

\item[{\sc III}]  The strength of each fiber is characterized by a {\it
critical   stress} $f_{c}(i)$: if at time $t$  the i-th fiber
local stress   $f_i(t)$ is greater than a critical stress
$f_{c}(i)$ the fiber   cracks, and the  local force falls to zero
at time $t+1$. Further,   we assume that in this process some
energy $\epsilon _i$,   proportional to the square of local force
$f_i$, is released.   For sake of simplicity we will assume
$\epsilon _i=0.5 \ f_i^2/Y$. The   critical stress $f_{c}(i)$ is a
realization of a random variable   that follows a normal
distribution of mean $<f_{c}>$ and variance   $KT_d$:
\begin{equation}   f_c(i)\sim N_d (<f_{c}>,KT_d)
\end{equation}     We call $N_d$  the {\it disorder
noise}.

 \item[{\sc IV}]  Each fiber is subjected to an additive
random stress $\Delta   f_i(t)$, which follows a zero mean normal
distribution of variance   $KT$:

\begin{equation}
\Delta f_i(t)\sim N_T(0,KT)
\end{equation}

 being $t$ the time. We call $N_T$ the {\it thermal noise}. We assume that $%
 \Delta f_i(t)$ is a white random process,which is  independent
in   each fiber, i.e.  the correlation function $E\left[ \Delta
  f_i(t_1)\cdot \Delta f_j(t_2)\right] =0$ if $t_1\neq t_2$ or
  $i\neq j$. A time step of the model corresponds to the
  application  of the  noise to all the fibers.
  \cite{noise-comment}

\end{itemize}

    The first three items are
those used in the standard formulation   of the fiber bundle
model. A  new feature,  which is similar to a   thermal activation
process, is introduced in [{\sc IV}] to explain the   dependence
of the failure time on a constant applied stress.       The model
has the following properties. We see from [{\sc I}] that   there
exists a long-range interaction between the fibers, that is, under
the specified boundary conditions, the breaking of some   fibers
produces an increase of the local force on the other ones.
 Indeed if a number
$n(t)$ of fibers is broken at time $t$ then  the local   force on
each of remaining fibers is:
\begin{equation}   f_i(t)=\frac { f_o \ N }{N-n(t)}+\Delta f_i(t)
\label{fdelta}   \end{equation} where $f_o=F/N$ is the initial
force per fiber.
 Procedure [{\sc II}] specifies the elastic behaviour of
fibers. Procedure   [{\sc III}] models the heterogeneity of the
medium: if there is no disorder noise    (i.e. $KT_d=0$) all the
fibers are strictly equal, and if there is no     thermal
activation (i.e. $KT=0$) the  medium breaks down instantaneously
when $F=N f^{(c)}$. Conversely, if $KT_d>0$ fibers begin to crack
before the final break-down of the media. The assumption that all
the disorder in the model appears in the strength distribution
rather
      than in the elastic constants may be argued by noticing
that the        effective elastic constant of a single fiber is
essentially the   average of the local elastic constant along the
fiber, while the         strength is determined by its weakest
point \cite{hemmer}.         A thermal activation process is
introduced in [{\sc IV}] to explain   the dependence of failure on
time; while [{\sc I-III}] are standard   procedures, [{\sc IV}] is
new with respect to DFBM.

  In this paper we want to study
   the failure time
$\tau $ of the sample as a function of $f, \ KT $ and $KT_d$.

\section{Some analytical results}

\subsection{The energy barrier}

   Eq.\ref{const} has been obtained by   Pomeau  by considering that the final
   fracture of a material is produced by a thermally activated
   nucleation process of microcracks.
   The energy barrier of this processes has been  estimated by using
    the Griffith criterion \cite{Lawn} for the stability of a
fracture  in  an elastic medium.
  In this section we
want to construct a similar   criterion   for the bundle by using
the same procedure used in \cite{pomeau,Lawn}. More specifically
we compute the energy barrier that the system has to overcome in
order  to break the bundle when $n$ fibers are broken and a total
constant force $F$ is applied. We begin by noticing that the
strain energy in the bundle with a constant tensile force is $U_E=
F^2 \lambda/2$ where $\lambda=1/[Y \ (N-n)]$ is the bundle
compliance. The change in
  the potential energy due to the damage extension   is
$\Delta U_A=-F^2 \Delta \lambda$. Thus the   mechanical energy is
    $U_M=U_A+U_E=- F^2 /[2 \ Y \ (N-n)]+ F^2 /(2\ Y \ N) $.
    The energy $U_f$ used to break $n$ fibers can  be computed
  by considering  the energy needed by $n$ fibers to reach the
  elongation corresponding to $f_c$, that is   $U_f= f_c^2
/( 2 \ Y ) n$.    Thus the total   energy ( for  Y=1) is
\begin{equation}   E_{tot}={1 \over 2}    \left[ {F^2
\over  N}-{F^2 \over N-n}+ f_c^2  n \right]   \label{Etot}
\end{equation}
 At $n=n_c=N- F/f_c$ this energy has a maximum:
\begin{equation}   E_{max}= {f_c^2
N \over 2} \left({F \over N f_c} -1\right)^2   \label{Emin}
\end{equation}
This means that for $n>n_c$ the force per bond is larger than
$f_c$. Thus the bundle is unstable and will break. Following
Pomeau's activation model the minimum energy per fiber $E_{max}/N$
has to be compared to $KT$ to estimate the lifetime of the bundle.
Thus the lifetime of the sample has  a different functional
dependence than eq.\ref{const}:

\begin{equation}   \tau = \tau_o \exp \left[ {f_c^2 \over 2 KT} \left({f
 \over  f_c} -1 \right)^2 \right]   \label{tau-activ}   \end{equation}

 where
$f=F/N$. Comparing  eq.\ref{const} with eq.\ref{tau-activ} we see
 that   the functional dependence of
$\tau$ on $f$ is different for the bundle (eq.\ref{tau-activ}) and
for a solid (eq.\ref{const}). This  difference is obviously due to
the different geometry of the tensile force redistribution in the
bundle and in the solid. In section 4 we will show, using
numerical data, that eq.\ref{tau-activ} well describes the
behaviour of the lifetime of the bundle, measured in numerical
simulations. However eq.\ref{tau-activ} can be obtained by a
statistical analysis of the DFBM with noise. A similar analysis
has been performed by Roux \cite{roux} and in the next section we
extend his analysis.

\subsection{Statistical analysis}
 We describe here only the main results. The details of the
 calculation can be found in the appendix.

\subsubsection{The lifetime of the homogeneous fiber bundle}

We first consider the homogeneous case where the threshold force
$f_c \equiv 1$ is the same for all of the fibers.
 Let $P(\eta) = \int _\eta ^\infty {\frac{1}{\sqrt{2\pi KT}}
  \exp \left( -\frac{x^2}{2KT} \right) dx}$
   the probability that  the thermal noise has an amplitude
   larger than $\eta$.
The mean force present on each fiber is $f = f_o \frac{N}{N-n}$.


It can be proved  that in the limit $N\rightarrow \infty$ the
ratio $\frac{n}{N}$ becomes independent of $N$ (see appendix).
Thus we can introduce the function $\phi(t)$ (fraction of broken
fibers at time $t$) and the complementary function $\phi_c(t)$,
defined as:
\begin{equation}
    \phi(t) = \lim_{N \rightarrow \infty}{\frac{n}{N}}
\end{equation}
\begin{equation}
    \phi_c(t) = 1-\phi(t)
\end{equation}
In the appendix we show that one can write the following equation
for $\phi(t)$:

\begin{equation}
    \phi(t) = \int_0^t \phi_c(\zeta)
     \cdot P\left( 1-\frac{f_o}{\phi_c(\zeta)} \right) d\zeta
\end{equation}
that is:
\begin{equation}
    \frac{d\phi(t)}{dt} = \phi_c(t) \cdot P\left( 1-\frac{f_o}{\phi_c(t)} \right)
    \label{Dphi}
\end{equation}

The failure time $\tau$ is defined as the time at which all  the
fibers are broken. In the appendix we show that in the limit
$\phi<<1$ and $2\sqrt{2 \ KT} < (1-f_o)$ it is possible to write
an approximated solution of
 eq.\ref{Dphi} with a normal distribution of the thermal noise.
 This solution is:

\begin{equation}
\tau  \simeq \frac{\sqrt{2\pi KT}}{f_o} \exp
\left(\frac{(1-f_o)^2}{2KT}\right)  \label{tau-stat}
\end{equation}

and

\begin{equation}
\phi(t) \simeq - \frac{1}{C} \ln \left(\frac{\tau-t}{\tau}\right)
\label{eq:phi} \end{equation}

where

$$ C= \frac{(1-f_o)f_o}{KT}. $$

Notice that, except for logarithmic corrections,  the main
dependence of $\tau$ on $f_o$ and of $KT$ in eq.\ref{tau-activ}
and in eq.\ref{tau-stat} is the same.





\subsubsection{The lifetime of the disordered fiber bundle}

In the disordered case the breakdown threshold $f_c$ is a random
variable. In order to find an analytical solution of $\tau$ as a
function of $f$,  one has to make several  approximations, which
are described in the appendix.  When $f_c$ is normally distributed
with mean $<f_c>=1$ and variance $KT_d=\sigma_o^2/2$, one finds
that in the limit of $\phi<<1$, $2\sqrt{2 \ KT} < (1-f_o)$ and
$2\sqrt{2 \ KT_d} < (1-f_o)$ a good  approximation of the failure
time is:

\begin{equation}
\tau \simeq \tau_o \exp \left(\frac{(1-f_o)^2}{  2 \
KT_{eff}}\right) \label{tau-disorder}
 \end{equation}
with

\begin{equation}
  KT_{eff}\simeq { KT \over \ \left( 1-\frac{\sqrt{\pi} \ \sigma_o}{2(1-f_o)} \right)^2 }
  \label{eq:Teff}
\end{equation}

and

\begin{equation}
 \tau_o=
 \frac{2 \ \sqrt{2\pi KT}}{(f_o-\sqrt{\pi} \  \sigma_o) \
 \left[1+\exp\right(-\frac {\sqrt{\pi}  \
\sigma_o \ (1-f_o)} {KT}\left) \right]}\label{eq:tauo}
\end{equation}
\medskip

One notices that for $KT_d\ne 0$ there is an effective
temperature, which is an increasing function of the disorder
variance $KT_d$. This important observation will be numerically
checked in sect.4.

Another interpretation of eq.\ref{tau-disorder} and
eq.\ref{eq:Teff} is that the disorder changes the critical force
from $f_c=1$ (in the ordered case) to a smaller value
($1-\sqrt{\pi} \sigma_o/2 $)  in the disordered case. We consider
that the interpretation in terms of $KT_{eff}$ is the best one.
Indeed one can show (see \cite{roux})that if the dynamics would be
dominated by the fibers with thresholds in the tails of $G_o(f_c)$
than $KT_{eff}=KT+KT_d$. The interpretation given in
eq.\ref{eq:Teff} stresses the difference between a tail dominated
dynamics and a dynamics dominated by the fibers with
$f_c>f_{min}$.

\section{Numerical results}

In this section we check the analytical results obtained in sect.3
on a numerical simulation on the fiber bundle.
 In the following we will assume without
loss  of generality $f_c=1$ and $N=1000$. We have checked that the
results do not change considerably up to $N=10^6$. For each values
of the parameters $KT$, $F$, $KT_d$ we have repeated the
experiments at least 10 times in order to estimate the scattering
of the results in different realizations.

\subsection{Failure time at constant force}

\subsubsection{The homogeneous case}

\begin{figure}
\centerline{ {\bf (a)} \hspace{7cm} {\bf (b)} }
 \centerline{ {\epsfysize=0.5\linewidth \epsffile{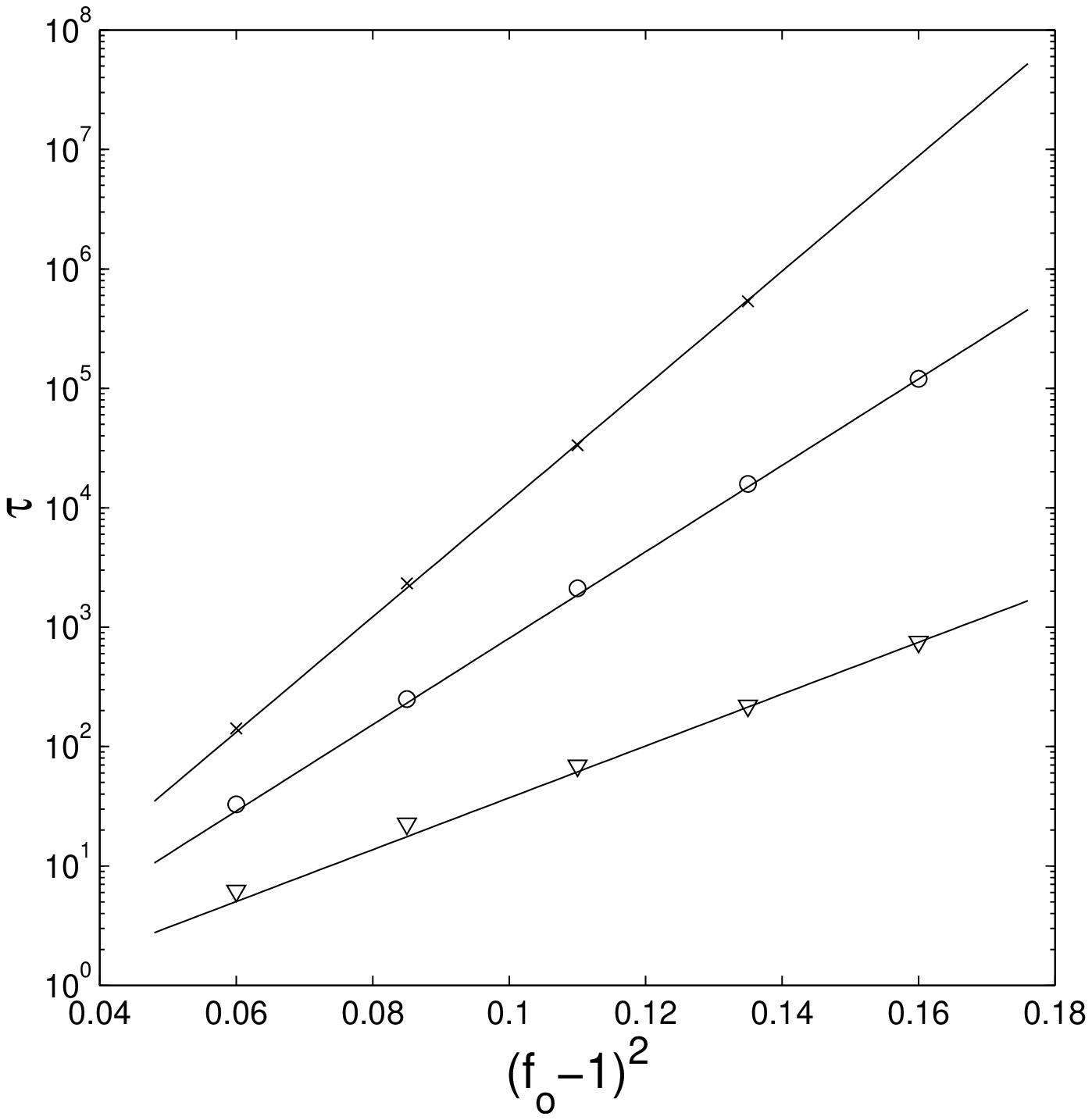}}
 {\epsfysize=0.5\linewidth \epsffile{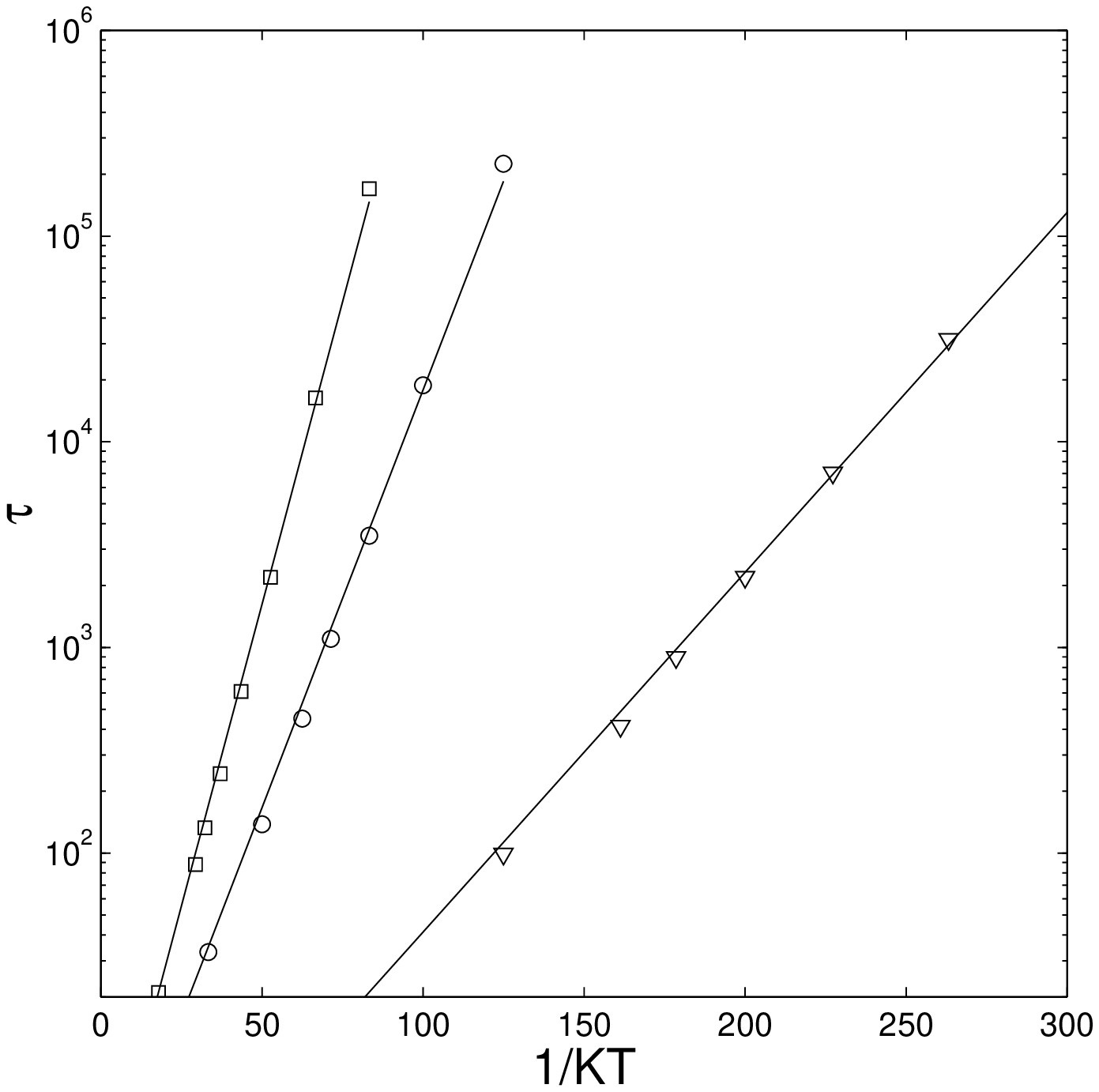}} }
  {\bf (c)} \\
  \centerline{\epsfysize=0.5\linewidth \epsffile{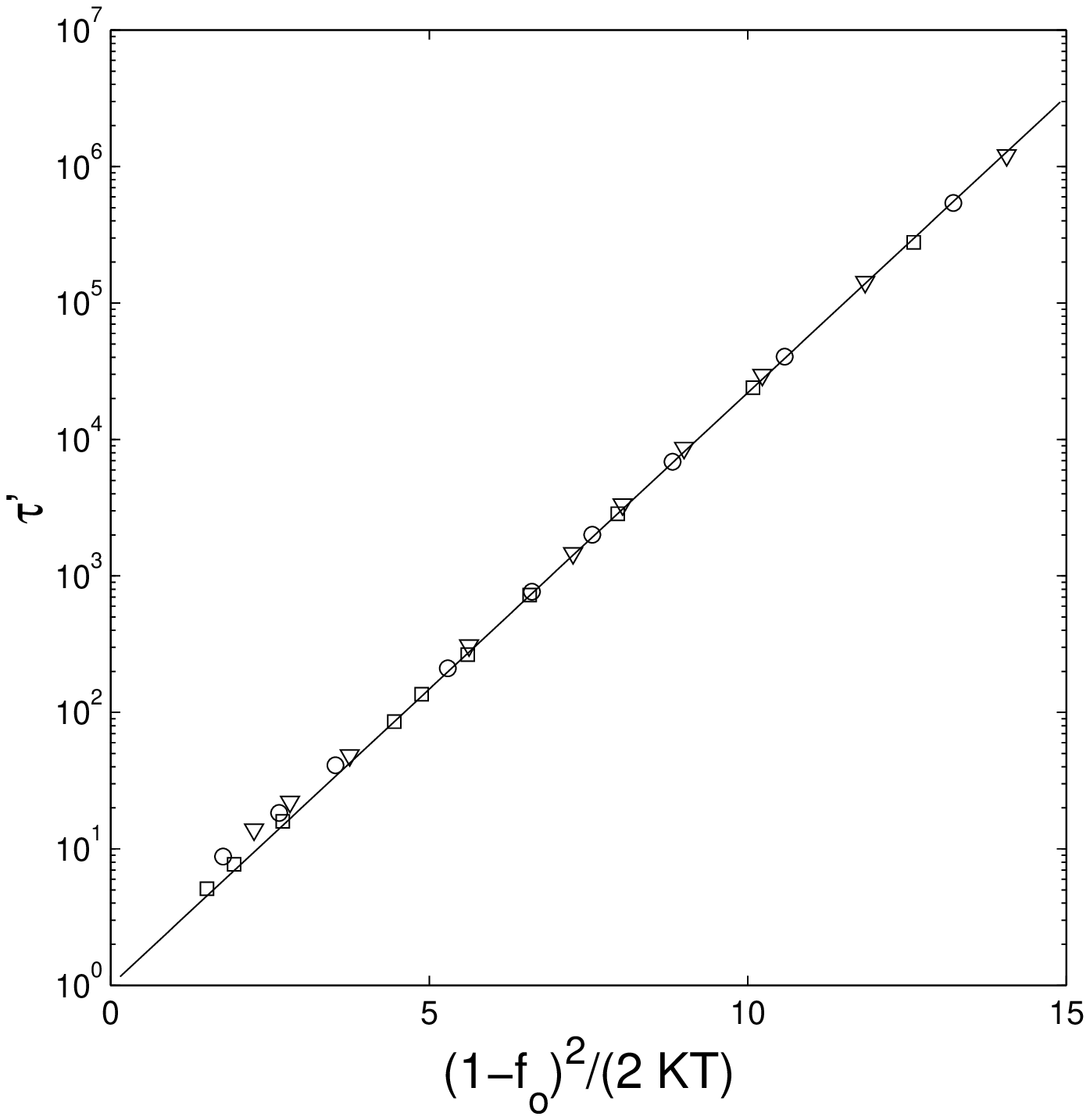}}
  \caption {Homogeneous net ($KT_d=0$)  in a creep test. (a) Failure  time $\tau $ as a
function of
  $(1-f_o)^2$ for several values of the thermal noise variance $KT$ ($(\times),KT=0.0045$, $%
  ({\circ}),KT=0.006$, $(\triangledown),KT=0.01$). The continuous lines
  are  the fits with eq.\ref{tauvf} (b) $ \tau$ as a function of
   $1/KT$ for several values of imposed force
   $f\cdot N. \ \ (\square) \ f_o=0.45$; $(\circ)  \ f_o=0.54$; $(\triangledown) \ f_o=0.7$.
    The continuous lines are  the fits with
  eq.\ref{tauvkt}. (c) $\tau'= \tau \ f \ (2 \pi \ KT)^{-1/2}$ as
  function of $(1-f)^2/(2 KT)$. The same symbols of (b) are used.
  The continuous line is obtained from  eq.\ref{tau-stat}.
  } \label{lifetime}
  \end{figure}

In order to study the behaviour of failure time $\tau $, we
started by imposing a force $F(t)=F$ constant in time (creep
test). If disorder noise and thermal noise are both zero (i.e.
 $KT= KT_d = 0$), the system  either breaks at time $\tau =0$ only if
  $F\geq Nf_c$ or it  does not break at all otherwise.
  Conversely, if thermal noise is not zero (i.e. $KT \neq 0$),
we   observe (fig. \ref{lifetime} a) that failure time $\tau $ has
an exponential dependence on
  $(1-f_o)^2$  for any fixed value of
$KT$,   specifically:
 \begin{equation}
 \tau \sim \exp \left[
\alpha (1-f_o) ^2 \right]    \label{tauvf}
\end{equation}
 where $\alpha$ is a fitting parameter, which is a function of $KT$.
 Furthermore    at constant
$f_o$ (see fig. \ref{lifetime}  b), the   failure time  $\tau $
depends on $KT$ as follows:
 \begin{equation}   \tau \sim \tau_o
\exp \left( \frac A {KT} \right) \label{tauvkt}
\end{equation}   where $A$ is a function of $f$. We notice that these results are   similar to
those  predicted by   the  activation model  discussed   in
section 3 and that the lifetime is well described by
eq.\ref{tau-activ}. More precisely one can check the predictions
of eq.\ref{tau-stat} by plotting in fig.\ref{lifetime}c
$\log({\tau \ f_o \over \sqrt{ 2\pi KT}} )$ as a function of
$(1-f_o)^2 \over  2 KT$. We see that all data collapse on a single
straight line  of slope $1$, verifying in this way  the hypotheses
made to obtain eq.\ref{tau-stat}. In fig.\ref{Fig:phi} we plot the
time evolution of $\Phi$ as a function of time. The continuous
line is the numerical solution of eq.\ref{Dphi}. The circles
correspond to the mean values  obtained in  several direct
numerical simulations of the DFBM.  In the inset we compare the
result of the direct numerical  simulation  with the approximated
solution (eq.\ref{eq:phi}) of eq.\ref{Dphi}.

    Fig.\ref{lifetime} and fig.\ref{Fig:phi} show that the
hypotheses made to obtain eq.\ref{tau-stat} and eq.\ref{eq:phi}
are well verified by the numerical integration of the model.

\subsubsection{The heterogeneous case}

\begin{figure}
 \centerline{\epsfysize=0.5\linewidth \epsffile{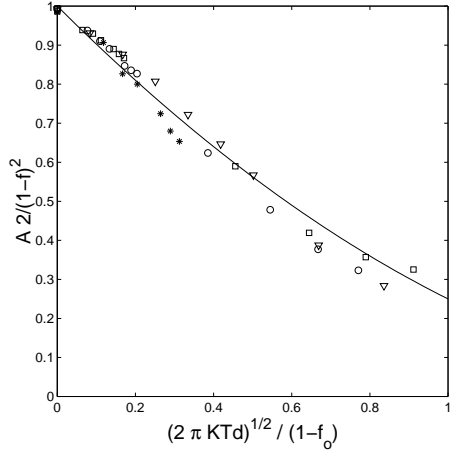}}
  \caption { Dependence of  $2 A/(1-f_o)^2$ ($A$ is defined in
  equation \ref{eq:A}) as a function of $\sqrt{2 \pi  \ KT_d}/(1-f_o)$ for different values
 of $f_o$: $(\square) \  f_o=0.45;(\circ) \ f_o=0.54;(\ast)  \ f_o=0.7$. The
  continuous line is the theoretical prediction  obtained from  eq.\ref{eq:A}.
   The
 symbols
($\triangledown$) correspond to a numerical simulation done with
uniform distribution at $f_o=0.54$. In this case the x axis is
$0.8 D/(1-f_o)$( see appendix B.2). } \label{fig:Teff}
\end{figure}

  We are now interested in  studying  the dependence of the
failure time   $\tau $ on disorder noise. To this aim, we have
done simulations   keeping $KT_d$  fixed at constant (non-zero)
values.
 In fig.\ref{Fig:time-disordine} we plot the
dependence of $\tau/\sqrt{2\pi KT}$, at $f_o=0.54$, as a function
of $1/KT$ for different values of $KT_d$. We notice that for the
same values of $KT$, $\tau$ decreases by several order of
magnitude by increasing the disorder noise. Most importantly, we
see that for any values of $KT_d$ the leading dependence of $\tau$
on $KT$ is the same of the ordered case (eq.\ref{tauvkt}), but the
value of $A$ and $\tau_o$ depend on $KT_d$. Specifically,
increasing $KT_d$ the value of $A$ decreases  whereas $\tau_o$
increases. Using  eq.\ref{tau-disorder} and eq.\ref{eq:Teff} one
obtains that $A$ in the heterogeneous case is:
\begin{equation}
  A={ 0.5  \ (1-f_o)^2 \ \left( 1-\frac{\sqrt{\pi} \ \sigma_o}{2(1-f_o)} \right)^2 }
  \label{eq:A}
\end{equation}
 From the best fits of the numerical data of
fig.\ref{Fig:time-disordine} one can measure the values of $A$  as
a function of $f$ and $KT_d$. The prediction of eq.\ref{eq:A} can
be accurately checked by plotting $2 \ A/(1-f_o)^2$ as a function
of $\sqrt{2\pi KT_d}/(1-f_o)$. In fig.\ref{fig:Teff} we clearly
see that the numerical values of $2 A /(1-f_o)^2$, obtained for
different values of $f_o$, collapse  on the same curve, which
agrees quite well with the theoretical prediction, continuous
curve, obtained using eq.\ref{eq:A}.

\begin{figure}
 \centerline{\epsfysize=0.5\linewidth \epsffile{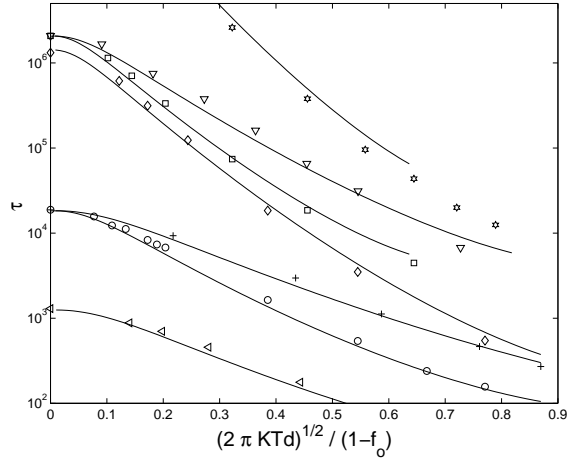}}
  \caption {
 The lifetime $\tau $ is plotted as a function of the  disorder
  noise variance $KT_d$ for several values of  $KT$ and of $f_o$ and different threshold
  distributions.
  The continuous  lines represent predictions of
  eq.\ref{tau-disorder} for the Gaussian distribution and eq.\ref{e:tau-uniforme}
for the uniform distribution. Discrete points are the results of
numerical measures.  For the Gaussian distribution:
   $(\circ) \  KT=0.01, f_o=0.54 ; (\square) \  KT=0.01, f_o=0.45 ; (\diamond)  \  KT=0.007, f_o=0.54 ;
   (\vartriangleleft)  \ KT=0.01, f_o=0.6;  (\star)  \ KT=0.0073, f_o=0.45$. For the
    uniform distribution $(+) \  KT=0.01, f_o=0.54 ;
(\bigtriangledown)  \ KT=0.01, f_o=0.45$.
    } \label{Fig:tauvktd}
  \end{figure}

The dependence of $\tau$ on $KT_d$ is plotted in
fig.\ref{Fig:tauvktd} for several values of $f$ and $KT$. The
continuous lines are the predictions of eq.\ref{tau-disorder}. We
see  that the main dependence of $\tau$ on $KT_d$ is very well
described by eq.\ref{tau-disorder}. The deviation at large $KT_d$
are due to the approximations made to get the analytical
expression eq.\ref{tau-disorder}. This equation and  numerical
results show that $\tau_o$ increases as a function of $KT_d$.
Therefore the observed reduction of the sample lifetime for
increasing $KT_d$ is mainly due to a decrease of $A$. In looking
at fig.\ref{Fig:time-disordine} and from eq.\ref{tau-disorder} we
also notice that the derivative of $\tau$ with respect to $KT$
decreases as a function of $KT_d$. Summarizing,  the following
conclusions can be extracted from the numerical and the analytical
results:

\begin{itemize}

 \item[B1] The failure time $\tau $ decreases as the disorder
  noise variance $KT_d$ increases,
  that is the more
the    medium  is heterogeneous, the smaller the failure time  is,
see fig.\ref{Fig:time-disordine}.

  \item[B2]  As the disorder
noise variance $KT_d$ increases, the derivative of $\tau $  with
respect to  $KT$ decreases, that is failure time $\tau $ becomes
less sensitive to the effective value of the thermal noise
variance, as shown in fig.\ref{Fig:time-disordine}.
\end{itemize}

  Thus the role of the spatial disorder is
that of multiplying $KT$ by a factor, which is an increasing
function of $KT_d$.
 This is
equivalent to say, as we already mentioned in section 3, that an
approximation of $KT_{eff}$ is given by eq.\ref{eq:Teff}.
Furthermore fig.\ref{Fig:time-disordine} and fig.\ref{Fig:tauvktd}
show that the disorder reduces the dependence of $\tau $ on the
variation of  the temperature $T$.

\subsection{Comparison with experimental results}

These  results allow us to give a reasonable  explanation of  the
experimental observations reported in the introduction. In fact,
experiments show that the lifetime $\tau$ of very heterogenous
materials is independent of $T$ \cite{noi}, while the lifetime
$\tau$ of quite homogenous materials as 2D-crystals  depends on
temperature $T$\cite{pauchard}. This is consistent with the
numerical results of our model. As shown in
fig.\ref{Fig:time-disordine} when $KT_d$ is small as in
2D-crystals  the dependence of $\tau $ on $KT$ remains important,
whereas this dependence becomes negligible when $KT_d$ is huge
(like in chip-board panel wood and fiberglass). In any case
disorder noise induces an effective temperature much larger than
the thermodynamic one $T$.

\subsection{Failure time distribution}

\begin{figure}
 \centerline{\epsfysize=0.7\linewidth \epsffile{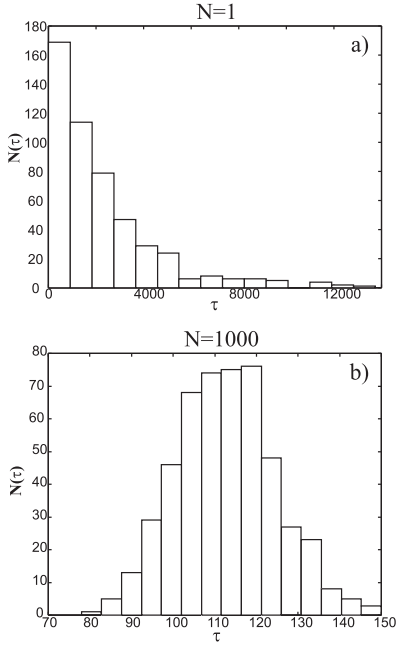}}
  \caption {Lifetime distribution for a single fiber (a) and (b)
  for $N=1000$.  $f_o=0.54$ and $KT=0.021$ in both cases}
  \label{time-distribution}
  \end{figure}

 We have
studied the distribution of failure time $\tau$   when the
parameters $KT$ and $F$ are constants, in the case of a   single
fiber and in the case of a net of many fibers. Clearly, in the
   case of a single fiber the probability of failure is
constant in time,     so that $\tau$ follows a Poisson's law
fig.\ref{time-distribution}. Conversely, in the case      of a net
of fibers one sees that $\tau$ is normally distributed.
 This result is
coherent with experimental results, which show that in highly
disordered materials and large aspect ratio  the sample lifetime
distribution is Gaussian.

\section{Generalization to a time dependent imposed force}

 In ref.\cite{noi} we proposed to generalize eq.\ref{tau-activ}
such that it may be applied   to any time dependent pressure; a
similar procedure is proposed here. Suppose that $\frac 1{\tau
(F)}$ is the damage density per unit time of the medium subjected
to a force $F$. For an arbitrary  load $F(t)$ the total damage
density at time $t$ is:
\begin{equation}   I(t) = \int_0^t \frac 1{\tau _0}\exp \left[
  -{(1-F(t')/N)^2 \over KT_{eff}}\right] dt' \label{integral-eq}
\end{equation}    thus the certitude of failure is
attained when $I(t) = 1$. Notice  that this yields
eq.\ref{tau-activ} if $F(t) = F$.\\ In order to   verify whether
this equation holds, we have done simulations imposing   $F(t)$
with different dependencies on time.

\begin{figure}
   \centerline{\epsfysize=0.5\linewidth \epsffile{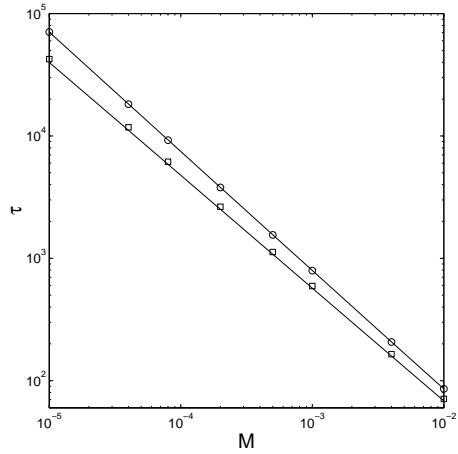}}
 \caption {Failure time $\tau $ as a function of the slope  $M$ in the case
  of a linear (in time) imposed force $F=M \ t$.
  The theoretical prediction of eq.\ref{eq:linear}(continuous
  line) confirms the results obtained in numerical simulations. In
  these simulations: $N=1000$, $KT_d=0$, $(\circ) \ KT=0.005$ and
  $(\square) \ KT=0.02$} \label{time-dep-force}
  \end{figure}

As an example we consider  a linear dependence on time of $F$:
$F=M \ t$. Inserting this linear dependence in
eq.\ref{integral-eq} we can approximately compute the dependence
of $\tau$ on the slope $M$ getting for $M<<\sqrt{2 \  \ KT \over \
\pi }$:
\begin{equation}
 \tau= \frac{ \ \left(1 -  \sqrt{2 \ KT } \ \ \delta_M \right)}{M \ \left(1+ {0.5 \ \sqrt{2 \ KT}
 \ / \  \delta_M }\right) }
 \label{eq:linear}
 \end{equation}
 where
 $$ \delta_M = \ln \left( \sqrt{2 \ KT \over \pi }  \  \frac{1}{M} \right) $$
¨
In fig.\ref{time-dep-force} we plot $\tau$ as a function of $M$ at
$KT=0.01$.  Circles are the computed values and
 the continuous line  is  the prediction of eq.
\ref{eq:linear}. The agreement between theoretical prediction  and
numerical simulations is quite good.

\begin{figure}
  \centerline{\epsfysize=0.8\linewidth \epsffile{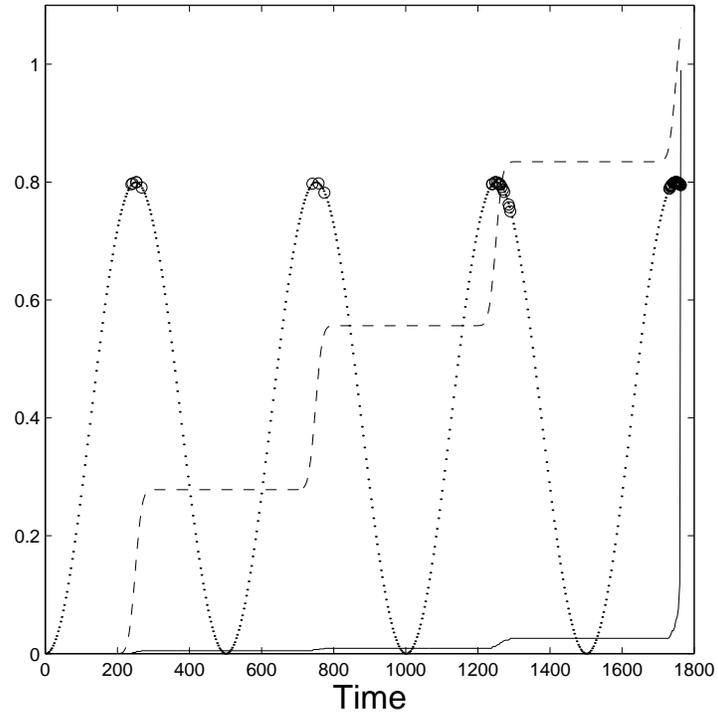}}
 \caption { A sinusoidal force (dotted line) is applied to the bundle.
 The dashed line represents the numerical integration
 of the integral in eq.\ref{integral-eq}. The circles indicate the
 occurrence of breaking events.
 Finally the continuous  line is the fraction of broken fiber $\phi$
as measured by the direct numerical simulation of the fiber
bundle.
 The time predicted by eq.\ref{integral-eq} perfectly agrees with the
that of the direct numerical simulation. In this simulation
$N=1000$, $KT_d=0$, $ \ KT=0.003$} \label{sinusoidal}
  \end{figure}

Further we apply to the bundle a sinusoidal force as shown in
fig.\ref{sinusoidal}. In this figure the dotted line is the
external applied force and the dashed line represents the
numerical integration of the integral in eq.\ref{integral-eq}. The
circles indicate the occurrence of breaking events. Finally the
dotted line is the fraction of broken fiber $\phi$ as measured by
the direct numerical simulation of the fiber bundle. We clearly
see that the lifetime predicted by eq.\ref{integral-eq} perfectly
agrees with that of the direct numerical simulation. This shows
that eq.\ref{integral-eq} gives a good estimate of the breaking
time for a periodic forcing.

\section{Kaiser effect}

   In 1950 Kaiser discovered that the acoustic emission of a stressed
     metal sample is zero if the
applied stress is smaller than the  previously applied maximum
\cite{kaiser}. This effect, usually   called {\it Kaiser effect},
was also discovered in rock  materials\cite{kurita,kanagaua}, but
its existence was seriously   questioned for the Westerley granite
\cite{granito} and for   heterogeneous materials as wood and
fiberglass \cite{noieps}. In our simulation we call {\it event}
the simultaneous breaking of several fibers, and   {\it size of
the event} the number $s(t)$ of fibers which break.   The energy
$\epsilon $ associated to an event is the sum of the   energies
released by the fibers that crack, that is :
\begin{equation}   \epsilon =\frac {s(t)}{2 \ Y} \ \left( \frac
F{N-n(t)}\right) ^2   \label{epsilon}
\end{equation}
 The potential energy of a fiber that breaks is
  proportional to its accumulated elastic potential energy.
   If we assume that the acoustic energy is proportional to
 this  potential energy then $\epsilon$ may be compared to the
 acoustic energy measured in the experiments \cite{noiprl,noieps}.
The cumulative energy $E_a(t)$ is the sum of $\epsilon$ from $0$
to $t$.

\begin{figure}
   \centerline{\epsfysize=0.9\linewidth \epsffile{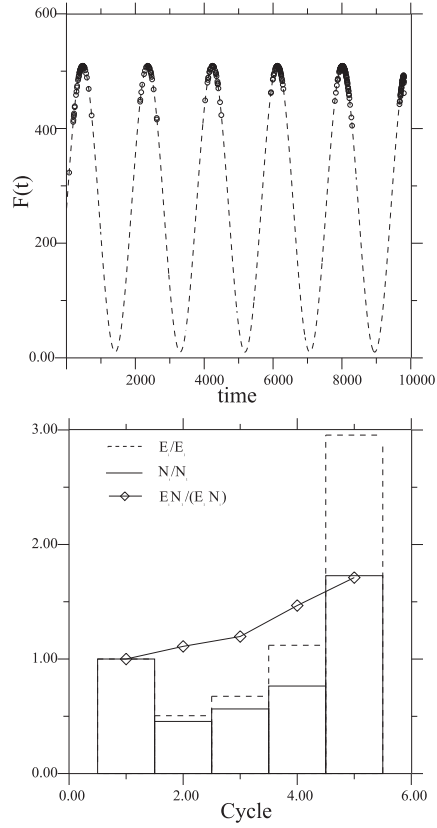}}
  \caption { (a) Imposed force $F=F_o[1-\cos (\frac{2\pi }{T_\omega
  }t)]$ as a function of time in a single realization; discrete
  points represent the occurrence of breaking events. (b) Number
  of events $N_i$ (continuous bars) and the energy $E_i$ (dashed
  bars) released during the i-th cycles, as functions of $i$. Both
  these quantities are normalized to the values of the first cycle.
  Discrete points represent the normalized mean energy
  $\frac{E_i/N_i}{E_1/N_1}$ of an event occurred during the i-th
  cycle. We observe that Kaiser effect is not valid, that is there
  are a large number of events after the first cycle.
  The parameters of this simulation are $N=1000$ and $KT=9\cdot 10^{-3}$ $%
  KT_d=1.5\cdot 10^{-2}$. }
  \label{kaiser}
  \end{figure}

In   order to check the behaviour of our model with respect to the
Kaiser effect we have imposed cyclic forces $F=F_0[1-\cos
(\frac{2\pi }{T_\omega }t)]$ to the net (the continuous line in
fig. \ref{kaiser}a). If $KT=0$, i.e. in a classic DFBM, the Kaiser
effect is   strictly valid: no events appear after the first
period.   Conversely, if $KT>0$, events (spots on fig.
\ref{kaiser}a) are produced  after the first. In fig.
\ref{kaiser}b), the number of events $N_i$ (continuous bars) and
the energy $E_i$ (dashed bars) released during the i-th cycles are
plotted, normalized to the values of the first cycle, as a
function of the cycle number $i$. The released energy $E_2$ and
the number of events $\allowbreak N_2$ occurred during the second
cycle are not negligible as they represent about the $50\%$ of
$E_1$ and $N_1$. From the second cycle ($i\geq 2$), $N_i$ and
$E_i$ grow with $i$. During the fourth cycle the energy released
$E_4$ becomes bigger than the energy $E_1$released during the
first cycle. The number of events $N_i$ grows slower than the
released energy $E_i$ so that the average energy $E_i/N_i$
released by an avalanche during the i-th cycle grows as a function
of $i$. The behaviour of $E_i/N_i$ is represented by circles on
fig. \ref{kaiser}b). The rates of the $N_i$ and $E_i$ after the
first cycle depend on $KT$ and $KT_d$. The statistic of events is
very similar to experimental observations in heterogeneous
materials \cite{noieps}. In conclusion, we have shown that the
Kaiser effect is valid only if the thermal noise is negligible.
Conversely, if a thermal noise is present the Kaiser effect is not
observed.

\section{Conclusions and discussion}

To simulate the failure of heterogeneous materials, we used a DFBM
with   thermal noise.
  We observe that the breakdown of the system occurs   through  the coalescence
  of many microcracks which are driven  by the thermal
  fluctuations and which appear  at different
  times. Thus   the lifetime $\tau $
  of the DFBM with  thermal noise follows a law, which is similar to an activation
  processes.
The main goal of the simulation was to check the role of the
  disorder on the lifetime of the system.
We have shown both numerically and analytically that the most
important effect of the disorder is that of producing an effective
temperature of the system, which is larger than that of the
thermal noise. This effective temperature (eq.\ref{eq:Teff}) is
equal  to the thermal noise temperature multiplied by a factor,
which is an increasing function of the disorder variance. This
important result is independent on the specific distribution of
thresholds. In this sense we can say that the disorder {\it
"amplifies"} the thermal noise temperature. It is important to
stress that that thermal noise and disorder play different roles:
 indeed  it is the thermal noise, which drives the cracking process,
 while the disorder just enhances its effect.
 Two  important hypotheses have been done to get this result. The first
  is that the threshold
 distribution evolves as a function of the fraction of broken
 bonds. The second is that the tails of the distribution do not play an important role
 in the fracture process. To estimate  $\tau$, one has to consider only the part of the
 distribution containing about $90\%$ of the unbroken bonds.
 The effective temperature of eq.\ref{eq:Teff} is quite different
  from the result recently obtained by
 Roux\cite{roux}. He finds that the effective temperature is the sum
 of  $KT$ and $KT_d$. Indeed our result is not directly comparable with that
 of Roux because he estimates the time needed to cut the first fibers.
This time is negligible with respect to the total lifetime of the
bundle. Furthermore the number of bonds contained in the tails
does not affect the resistance of the system (for example  in the
case of the Gaussian only $2\%$ of the bonds have a threshold in
the tails).

\begin{figure}
\centerline{\epsfysize=0.5\linewidth \epsffile{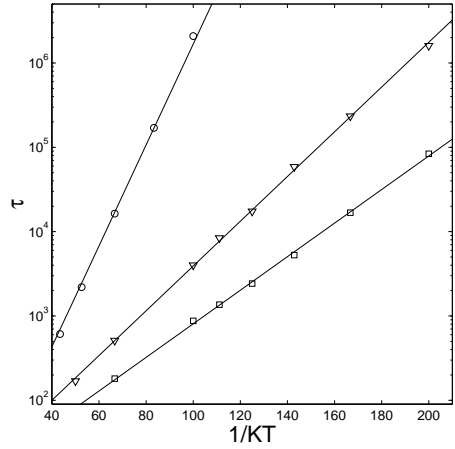}}
  \caption {  Heterogeneous bundle ($KT_d \ne 0$).
   Dependence of the lifetime $\tau$ on $1/KT$ at $f_o=0.45$.   The different
    symbols correspond to different values of $KT_d$:  $(\circ)  \  KT_d=0,
(\triangleleft) \ KT_d=0.02, \ (\square)  KT_d=0.04$
 Notice  that both
  $\tau $ and $\frac{d\tau }{ dKT}$ decrease as $KT_d$ increases,
  that is the more media become heterogeneous, the smaller is $\tau
  $ and the dependence on $KT$ of $\tau$. }
  \label{Fig:time-disordine}
  \end{figure}

Our results not only demonstrate that the disorder enhances the
role of thermal noise but eq.\ref{eq:Teff} and
fig.\ref{Fig:time-disordine} clearly show that when the  disorder
noise variance
  $KT_d$ increases, the {\it absolute} value of the derivative $ d\tau \over d \ KT$  decreases,
  that is the  failure time $\tau $ becomes less sensitive to variation of  the
  effective value of thermal noise (not zero as well).
 The conclusion,  that in an activation processes disorder noise
  induces an effective temperature larger than
 the thermodynamic one, has been reached for  other disordered systems such as foams
\cite{sollich}. This kind of disorder induced enhancement of the
thermodynamic temperature is an interesting result because it
could be be a quite general property of disordered systems, where
a thermal activated processes with long range correlation is
present.

It is interesting to compare the  results of our model  with the
recent experimental observations  on the lifetime of samples
submitted to a constant applied stress.
  It has been shown that in
      wood \cite{noi},
fiberglass \cite{noi}, gels \cite{bonn}
  and 2D-crystals \cite{pauchard}
  the dependence of the  lifetime on the applied stress is well predicted by
  the Pomeau model on the delayed fracture (see eq.\ref{const}).
  However we already mentioned in the introduction that in all of
  the experiments the estimated temperature was higher than the
  thermodynamic one and in ref.\cite{noi} we argued
  that the heterogeneity of the material may enhance the  role of
  temperature. The theoretical and numerical  results of this
  paper give  new insight to this argument.
  The fiber bundle and the Pomeau model have a different
  dependence on the applied load (compare eq.\ref{tau-activ} and
  eq.\ref{const}). This difference comes from the fact that the
  energy barrier is different in a real crack and in the bundle.
Thus we neglect this difference and we focus on the role played by
 the disorder. The Pomeau  model is based on the idea that
  failure of the sample is due to the thermal nucleation of one
  defect (microcrack) initially present inside the material
  \cite{pomeau}. In contrast, this model and experiments show that
  the failure is due to the coalescence of many microfractures
  appeared at different times. We think that a good theoretical
  model to describe the dynamic of the DFBM and various real
  materials submitted to an external force could be made transforming
  in a statistical model the one of Pomeau. In that model the
  nucleation of a microcrack is the elementary process, i.e. the
  sample break when a certain
  number of microcraks have nucleated. In this case, the parameters $Y$, $%
  \Gamma $ and $T_{eff}$ of eq.\ref{const}, become average parameters. The strong
  time-dependent fluctuations of the internal forces induced by the
  heterogeneity (defects, microcraks ...), can be considered  as a sort
  of noise so that, in our statistical model, $T_{eff}$ must depend on the
  thermal temperature but also on the disorder in the medium.
As we have already mentioned our numerical and analytical results
on the fiber bundle also show that the dependence on the lifetime
on temperature decreases when the disorder increases.
   These results
  could explain why in very disordered materials, such as wood and
  fiberglass, the lifetime $\tau $ seems to be independent of
    temperature while $\tau $ strongly depends on temperature in weak
  heterogeneous materials such as 2D-crystals.

The comparison of the lifetime of the bundle with those of real
materials merits a special comment. Indeed in the bundle we
consider as unit of the time a complete update of all the fibers.
One can assume that the characteristic time in a real material is
the time needed to redistribute the stress after that a
microfracture is occurred. This time, which is given in a first
approximation by the size of the sample divided by the sound
speed, is of the order of $10^{-5} s$ \cite{noi}. This means that
the largest observed  experimental time  correspond  to roughly
$10^{10}$ time steps of the bundle. This very large $\tau$ are
reached for very small $KT$ and $KT_d$ where our approximation
becomes more precise. Thus we may assume that the comparison of
this theoretical results with experiment could be rather
realistic.

  In the last part of the paper we have introduced the concept of density of
   damage per unit time
   finding  an equation (eq.\ref{integral-eq}), which predicts the lifetime
  $\tau $ of the DFBM for any time dependent imposed force. These
  numerical results agrees with experiments on wood and
  fiberglass\cite{noi}. Finally, we have shown that the validity of
  Kaiser effect is linked to the value of the thermal noise $KT$ and
  of the disorder noise $KT_d$.

\smallskip

{\bf Acknowledgement} Part of this work has been done within a
SOCRATES exchange program of the European Community between ENSL
and "Facolt\'a d'Ingegneria, Univerist\'a di Firenze, Italy ". One
us (R.S.) thanks "Le laboratoire de Physique de l'E.N.S.L." for
the very kind hospitality during his visit.

\bigskip
 $\clubsuit$ Actual
address: Ecole Centrale de Lyon, 69131 Ecully, France.

\newpage

 \centerline{ \bf {\Large APPENDIX}}

\bigskip

\begin{figure}
    \centerline{\epsfysize=0.5\linewidth \epsffile{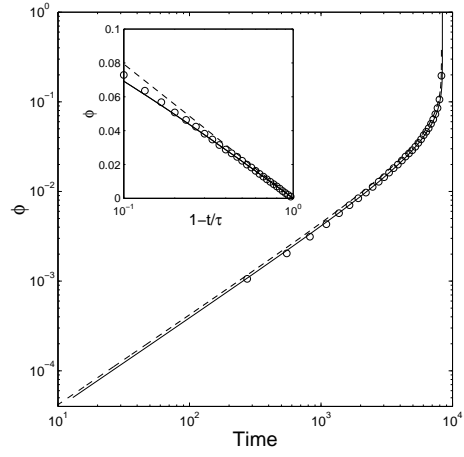}}
\caption{ Time evolution of $\phi$ at $f=0.6$,  $KT=0.008$ and
$KT_d=0$.  The continuous line  corresponds  to a solution
obtained by the numerical solution  of eq.\ref{Dphi}. The dashed
line is instead the approximated  solution  (eq.\ref{eq:phi}) of
eq.\ref{Dphi}. The symbols $(\circ)$ correspond to the results of
the direct numerical simulation of the DFBM. These data are
plotted in the inset as a function of $1-t/\tau$ in a semilog
scale.}
 \label{Fig:phi}
  \end{figure}

\centerline{\bf {\Large Analytical results}}
\appendix
{ \numberwithin{equation}{section}




\section{The homogeneous case}
Consider  the homogeneous case where the threshold force $f_c$ is
the same for all of the $N$ fibers composing  the bundle. Let
\begin{equation}
  P(\eta) = \int _\eta ^\infty G_T(x) dx=\int _\eta ^\infty {\frac{1}{\sqrt{2\pi kT}}
  \exp \left( -\frac{x^2}{2kT} \right) dx} \label{Peta}
\end{equation}
   be the probability that  the thermal noise has a
   fluctuation larger than $\eta$.
   If $n$ is the number of broken fibers at time $t$ then the
    force present on each fiber is $f = f_o \frac{N}{N-n}$.
    Thus the probability that a single fiber breaks in a time step is:
\begin{equation}
p_1 = P(f_c-f) = P \left( f_c-\frac{f_o N}{N-n} \right)
\end{equation}
The expected number of fibers that break in a time step is:
\begin{equation}
R_N = (N-n) \cdot p_1 = (N-n) \cdot P \left( f_c-\frac{f_o N}{N-n}
\right)
\end{equation}
Notice  that if this expression is smaller than one (i.e. $R_N<1$)
 one may think of $\left< t \right> = \frac{1}{R_N}$ as the expected
 time between two consecutive microcracks.

 These expressions  allow us to get several analytical results. Indeed it can be
 easily proved that as $N \rightarrow \infty$ the ratio $\frac{n}{N}$
  becomes independent of $N$:
\begin{eqnarray}
    \frac{n_t}{N} & = & \frac{1}{N} \sum_{i=1}^t R_N(i)\\
    & = & \frac{1}{N} \sum_{i=1}^t (N-n_i) \cdot P \left( f_c-\frac{f_o N}{N-n_i} \right) =\\
    & = & \sum_{i=1}^t\left( 1-\frac{n_i}{N} \right) \cdot P
    \left( f_c-\frac{f_o}{\left( 1-\frac{n_i}{N} \right)}\right)\label{eq:n/N}
\end{eqnarray}
We can introduce the function $\phi(t)$ (fraction of broken fibers at time $t$)
and the complementary function $\phi_c(t)$, defined as:
\begin{equation}
    \phi(t) = \lim_{N \rightarrow \infty}{\frac{n_t}{N}}
\end{equation}
\begin{equation}
    \phi_c(t) = 1-\phi(t)
\end{equation}
Turning the sum in eq.\ref{eq:n/N} into an integral one can write
the following equation for $\phi(t)$:
\begin{equation}
    \phi(t) = \int_0^t (1-\phi(\zeta)) \cdot P\left( f_c-\frac{f_o}{1-\phi(\zeta)} \right) d\zeta
\end{equation}
that is:
\begin{equation}
    \frac{d\phi(t)}{dt} = (1-\phi(t)) \cdot P\left( f_c-\frac{f_o}{1-\phi(t)} \right)
    \label{e:inizio}
\end{equation}

The failure time $\tau$ is defined as the time at which all of the fibers are broken.
It can be expressed in terms of $\phi(t)$ by the condition:
\begin{equation}
\phi(\tau) = 1
\end{equation}

In the homogeneous case we assume $f_c \equiv 1$ for all of the
fibers.
 Unfortunately no exact solution of \eqref{e:inizio} is known,
  so one has to make some approximations to get an  analytical result.
  The crucial point is that in the direct numerical simulation of DFBM one observes that
   $\phi \ll 1$ for most of the time: this allows us
   to write $\frac{1}{1-\phi} \simeq 1+\phi$ and to approximate $P(\eta)$
    using an asymptotic development:
\begin{equation}
P(\eta) = \frac{1}{2}
\left(1-\text{erf}(\frac{\eta}{\sqrt{2kT}})\right)
 \simeq \sqrt{\frac {kT}{2\pi}} \frac{1}{\eta} \exp\left(-\frac{\eta^2}{2kT}\right)
\notag
\end{equation}
Thus one can write:
\begin{eqnarray}
\frac{d\phi}{dt}  & \simeq &  \sqrt{\frac {kT}{2\pi}}
\frac{1-\phi} {\left( 1-\frac{f_o}{1-\phi} \right)} \exp\left(
-\frac{\left( 1-\frac{f_o}{1-\phi}
 \right)^2}{2kT} \right) \notag \\
& \simeq & \sqrt{\frac {kT}{2\pi}} \frac{1-\phi}{1-f_o-f_o\phi}
 \exp\left( -\frac{\left( 1-f_o-f_o\phi\right)^2}{2kT} \right)
  \notag
\end{eqnarray}
being $f_o\phi \ll 1$, one may write:
\begin{eqnarray}
\frac{d\phi}{dt} & \simeq & \sqrt{\frac {kT}{2\pi}}
\frac{1}{1-f_o}
 \exp\left( -\frac{\left( 1-f_o\right)^2}{2kT} \right)
 \exp\left(\frac{2(1-f_o)f_o\phi}{2kT}\right)\\
& = & B \ \exp\left(C\phi\right)
\end{eqnarray}
being $B$ and $C$ defined as:
\begin{eqnarray}
B & = & \sqrt{\frac {kT}{2\pi}} \frac{1}{1-f_o}
 \exp\left( -\frac{\left( 1-f_o\right)^2}{2kT}\right)  \\
C & = & \frac{(1-f_o)f_o}{kT}
\end{eqnarray}
At this point one turns to the integral equation:
\begin{equation}
\int {\exp\left(-C\phi\right)} d\phi = B \int dt \notag
\end{equation}
 The last equation can be integrated:
\begin{equation}
\frac{\exp\left(-C\phi\right)}{C} = -B\ t+\frac{1}{C}
\label{e:integrata}
\end{equation}
where the initial condition is $\phi(0) = 0$.\\

We verified numerically that this equation describes
 the system behaviour for small $\phi$. Moreover one
  finds that as $\phi$ becomes big enough, say $\phi_{break} \simeq 0.1$,
  the system undergoes a cascade of microcracks until the final breakdown,
   which occurs after a little time. Eq.\eqref{e:integrata} seems to
    represent closely the system's behaviour for almost all the time.\\

Thus one can write the following estimation for the failure time
$\tau$:
\begin{equation}
\tau \simeq \frac{1}{C\ B} = \frac{\sqrt{2\pi kT}}{f_o} \exp
\left(\frac{(1-f_o)^2}{2kT}\right)  \label{e:tau}
\end{equation} This expression of $\tau$ agrees with both
numerical
simulations and theoretical results of \cite{roux}.\\

Using \eqref{e:tau} into \eqref{e:integrata} one finds:
\begin{equation}
\frac{\tau-t}{\tau}  =  \exp\left(-C \  \phi \right)
 \label{e:t_phi}
\end{equation}

\begin{eqnarray}
 \ln \left( \frac{\tau-t}{\tau} \right) & = & -C \ \phi  \notag
\end{eqnarray}
Finally one finds:
\begin{equation}
\phi(t) = - \frac{1}{C} \ln \left(\frac{\tau-t}{\tau}\right).
\label{e:phi} \end{equation} We verified (see fig.\ref{Fig:phi})
that these equations
 agree very well with numerical data, actually even for values of $\phi$ comparable to $1$.
 From eq.\ref{e:t_phi} one obtains that $t\simeq 0.95 \tau$   at  $\phi=  3 /
 C$. Thus the system undergoes a fast cascade of microcracks until
 the final break up for $\phi > \phi_{break}=3/C$:

\begin{equation}
\phi_{break} = \ \frac { 3 \ KT} {f_o \ (1-f_o)}\  = \frac { 3 \
(1-f_o) } { 2 \ f_o \ ( \ln{\tau}- 0.5 \ \ln { \frac {2 \pi KT }
{f_o^2} } ) }
 \label{e:phi_break}
 \end{equation}

This equation verifies  that  $\phi<<1$ for any value of $f_o$ if
$KT$ is small enough.

\section{The disordered case}

In the disordered case the breakdown threshold $f_c$ of fibers is
a random variable.  Let $G(f_c)$ be the distribution of
thresholds, that is $\int_\eta^\infty G(f_c) df_c$ is the fraction
of fiber with threshold  bigger than $\eta$. The crucial point is
that the distribution of $f_c$ evolves with time (i.e. $\phi$) as
fibers break, in such a way that:

\begin{equation}
\int_{0}^\infty G(f_c,\phi) df_c =1- \phi \label{eq:G}
\end{equation}

The expected fraction of fibers which break at a given time $t$
is:


\begin{eqnarray}
R_\phi &=& \int_{0}^\infty P\left(f_c-\frac{f_o}{1-\phi}\right)
G(f_c,\phi)\  df_c \label{eq:RN1}
\end{eqnarray}
We use  the definition of $P(\eta)$ (see eq.\ref{Peta})  to write:
\begin{eqnarray}
R_\phi &=& \int_{0}^\infty \int_{\left(\xi-f_o'\right)}^\infty \
G_T(x)   \ G(\xi,\phi) \ d\xi \  dx  \notag \\
 &=& \int_{0}^\infty
\int_{\xi}^\infty \ G_T(f_c-f_o')   \ G(\xi,\phi) \ d\xi \ df_c \
\label{eq:RN2}
\end{eqnarray}

where $f_o'=\frac{f_o}{1-\phi}$.

 The function $\psi(f_c,\xi)=G_T(f_c-f_o')
\ G(\xi,\phi)$ is integrated in the plane $(f_c,\xi)$ within the
region {\it Re} which is defined by the intervals $ \xi \leq
f_c<\infty$ and $0\leq \xi<\infty$. In eq.(\ref{eq:RN2}) one may
invert the  order of the integration by considering that the
intervals defining region {\it Re} are: $0 \leq \xi < f_c$ and $0
\leq f_c<\infty$. Therefore eq.(\ref{eq:RN2}) becomes:

\begin{eqnarray}
 R_\phi &=& \int_{0}^\infty
G_T\left(f_c-\frac{f_o}{1-\phi} \right) \ df_c \int_0^{f_c}
G(\xi,\phi) \ d\xi \ . \label{eq:RN}
\end{eqnarray}

 Thus in the disordered case one may turn eq.\eqref{e:inizio}
into:
\begin{equation}
\frac{d\phi}{dt} = \int_0^\infty
G_T\left(f_c-\frac{f_o}{1-\phi}\right) df_c \int_0^{f_c}
G(\xi,\phi)\  d\xi \label{e:disordine}
\end{equation}

with the condition imposed by eq.\ref{eq:G}. Notice that if the a
delta distribution with normalization \ref{eq:G} is chosen for $G$
then one gets from eq.\ref{e:disordine} the equation for the
homogeneous case eq.\eqref{e:inizio}.
 Equation eq. \ref{e:disordine}  is far too complex to be integrated
analytically. However
 one can greatly simplify this framework by assuming that
the dynamics is controlled by the weakest fibers which are the
 first to break.
 This hypothesis
can  been checked in a direct numerical simulation using for
$G_0(f_c)$ a  Gaussian and an uniform distribution centered at
$f_c=1$. The result of the numerical simulation are shown in
fig.\ref{verifica-Gauss} for the Gaussian distribution. We clearly
see that are the weakest fibers which break first and that the
distribution evolves just on the left side.

Starting from this numerical evidence of the behaviour  of
$G(f_c)$ one can find an approximated analytical solution of
eq.\ref{e:disordine}.

\subsection{Case of the normal distribution for thresholds}

Let the initial distribution $G_0(f_c)$ of thresholds be a normal
distribution of variance $kT_d=\sigma_o^2/2$
 and mean value $\left< f_c \right> = 1$.
In fig.\ref{verifica-Gauss} we see that
 ( for $2\sqrt{2KT}<(1-f_o)$ ) when $\phi$ increases the distribution is  deformed
on the left side, whereas the right side remains unchanged.  On a
first approximation one can take into account this behaviour by
splitting  the distribution at time $t$:

\begin{equation}
G(f_c) = \frac{1}{\sqrt{\pi} \sigma_o} \exp\left( -
\frac{(f_c-1)^2}{\sigma_o^2} \right) \ \ \ \ \text{for} \ \ f_c>1
\end{equation}

and

\begin{equation}
G(f_c) = \frac{1}{\sqrt{\pi} \sigma_o} \exp\left( -
\frac{(f_c-1)^2}{\sigma_d^2} \right) \ \ \ \ \text{for} \ \  f_c<1
\end{equation}

where $\sigma_d$ depends on $\phi$. The normalization condition
\ref{eq:G} yields  $\sigma_d=\sigma_o(1-2\phi)$, which roughly
agrees with the numerical data. Thus the integral in
eq.\ref{eq:RN} becomes

\begin{eqnarray}
R_\phi = \frac{1}{\sqrt{\pi}} \int_{0}^1  \exp\left( -
\frac{(f_c-\frac{f_o}{1-\phi})^2}{\sigma^2} \right)
\left(1-\text{erf}(\frac{1-f_c}{\sigma_d}) \right) \frac{\sigma_d
\ df_c}{ 2 \ \sigma_o \ \sigma} + \notag
\\
\frac{1}{\sqrt{\pi}} \int_1^\infty  \exp\left( -
\frac{(f_c-\frac{f_o}{1-\phi})^2}{\sigma^2} \right)
\left(\frac{\sigma_d}{\sigma_o}
+\text{erf}(\frac{f_c-1}{\sigma_o}) \right) \frac{df_c}{ 2 \
\sigma}
 \label{integral}
\end{eqnarray}

In order to find an analytical expression  for eq.\ref{integral}
one has
to consider that in the case of a Gaussian distribution $90 \% $
of the unbroken fibers have a threshold $f_c>f_{min}=(1-\sqrt{\pi}
\ \sigma_d/2)$.
 One can make the hypothesis that only these fibers, with a large threshold,
  play an
important role in supporting the applied load. The mathematical
counter part of this hypothesis is that  only the part of the
distribution with $f_c>f_{min}$ has to be taken into account. Thus
in  the first integral of  eq.\ref{integral} the lower integration
limit becomes $ f_{min}=(1-\sqrt{\pi} \ \sigma_d/2)= [1-\sqrt{\pi}
\ \sigma_o(1/2-\phi)]$ and we can use  $ \ \text{erf}(x)\simeq 2
x/\sqrt{\pi} \ $. After applying these approximations one gets:

\begin{eqnarray}
R_\phi = \frac{1}{\sqrt{\pi}} \int_{f_{min}}^{1+{\frac{\sqrt{\pi}
\ \sigma_o}{2}}}  \exp\left( -
\frac{(f_c-\frac{f_o}{1-\phi})^2}{\sigma^2} \right)
\left(\frac{f_c-f_{min}}{\sigma_o \ \sqrt{\pi} } \right) \frac{ \
df_c}{   \sigma} + \notag
\\
\frac{1}{\sqrt{\pi}} \int_{1+{\frac{\sqrt{\pi} \
\sigma_o}{2}}}^\infty \exp\left( -
\frac{(f_c-\frac{f_o}{1-\phi})^2}{\sigma^2} \right) \left( 1-\phi
\right) \frac{df_c}{ \sigma}
 \label{integral1}
\end{eqnarray}

Taking into account that,within the integration limits of the
first integral, $({f_c-f_{min}})$ is almost constant with respect
to the variation of   $ \ \exp\left( -
\frac{(f_c-\frac{f_o}{1-\phi})^2}{\sigma^2} \right)$, one can take
the mean value of ${f_c-f_{min}}$. Therefore one obtains the
following approximated result:
\begin{eqnarray}
R_\phi = \frac{1-\phi}{2} \  \left[2- \ \text{erf}\left(
\frac{f_{min}-\frac{f_o}{1-\phi}}{\sigma} \right)
 -  \ \text{erf}\left( \frac{ 1+{\frac{\sqrt{\pi}\sigma_o}{2}}
 -\frac{f_o}{1-\phi}}{\sigma} \right) \right]
 \label{integral3}
\end{eqnarray}

In this equation one can use the asymptotic approximation of the
error function $\text{erf}(x)\simeq 1-\exp(-x^2)/(x \sqrt{\pi})$.
Thus one gets the equation describing the time evolution of
$\phi$.

\begin{equation}
\frac{d\phi}{dt} \simeq  B_n(1-\phi)\exp\left(A_n\phi\right)
\end{equation}
where the coefficients $A_n$ and $B_n$ are the following:
\begin{eqnarray}
A_n =  \frac{(1-\frac{\sqrt{\pi}\ \sigma_o}{2}-f_o)(f_o-\sqrt{\pi}
\ \sigma_o)}{kT}
\end{eqnarray}

\begin{eqnarray}
B_n =  \sqrt{\frac{2\pi}{kT}} \  \frac{
(1+\exp(-\frac{\sqrt{\pi} \sigma_d(1-f)}{kT})}{2\ (1-\frac{\sqrt{\pi} \ \sigma_o}{2}-f_o)} \times \notag \\
 \notag \\ \exp\left( -\frac{\left( 1-\frac{\sqrt{\pi} \ \sigma_o}{2}-f_o\right)^2}{2kT}\right)
\end{eqnarray}

 Following the same procedure used in the homogeneous case, one
 finds that the failure time is:

\begin{eqnarray}
\tau \simeq \frac{2 \ \sqrt{2\pi kT}}{(f_o-\sqrt{\pi} \ \sigma_o)
 ( 1+\exp(-\frac{\sqrt{\pi} \sigma_d (1-f)}{kT}) )} \notag  \times
 \\   \notag  \\
 \exp \left[\frac{(1-f_o)^2} {2\ kT}\cdot\left( 1-\frac{\sqrt{\pi }\
\sigma_o }{2 \ (1-f_o)} \right)^2\right] \label{e:tau-normale}
\end{eqnarray}

Notice that for $\sigma_d\rightarrow 0$ one recover eq.\ref{e:tau}
for the life time in the homogeneous case. In section  (4) we have
seen that the $\tau$ predicted by eq.(\ref{e:tau-normale}) agrees
very well with the  results of the numerical simulations.

\subsection{The uniform distribution}
 The same calculation
can be done taking  for $G(f_c)$ in eq.\ref{eq:G} a uniform
distribution of mean value $<f_c>=1$ and width $D$. We find in
this case that:

\begin{eqnarray}
\tau \simeq \frac{\sqrt{2\pi kT} \ 2}{(f_o-0.8*D)
 ( 1+\exp(-\frac{0.8 \ D (1-f_o)}{kT}) )} \notag  \times
 \\   \notag  \\
 \exp \left[\frac{(1-f_o)^2} {2\ kT}\cdot\left( 1-\frac{0.8\ D }{2 \ (1-f_o)} \right)^2\right]
\label{e:tau-uniforme}
\end{eqnarray}

where the factor $0.8 D$ comes from the fact that in the case of
the uniform distribution $90 \%$ of the the unbroken bonds have
$f_c> f_{min}(1-0.8 \ D/2)$.

\section{Discussion}

One observes that in both cases of uniform and normal
distributions, the presence of a disorder on thresholds have the
same effect over the general behaviour of the system. In
particular one sees that disorder enhances the effect of thermal
noise, in that it decreases in a multiplicative way the exponents
in \ref{e:tau-normale} and \ref{e:tau-uniforme}.
 We stress that thermal noise and disorder play different roles:
 indeed  it is the thermal noise which drives the cracking process,
 while the disorder just enhances its effect.
 The effective temperature of eq.\ref{eq:Teff} is quite different
  from the result recently obtained by
 Roux. He finds that the effective temperature is the sum
 of  $KT$ and $KT_d$. Indeed our result is not directly comparable with that
 of Roux because he estimates the time needed to cut the first fibers.
To compute this time Roux  considers that the threshold
distribution is almost quenched. This is certainly correct at the
very beginning of the dynamics
   but it is not at longer times as we have shown in previous
  sections. In any case the time needed to cut the first fibers
  is negligible with respect to the total life time of the bundle
if $(1-f_o)>\sqrt{2 \ KT_d}$.
 Furthermore the number of bonds which
have the threshold in the tails of the distribution does not
affect the resistance of the system, for example in the case of
the Gaussian only $2\%$ of the bonds have the threshold in the
tail and in the case of a uniform distribution there are no tails.

}

  \newpage

  \bigskip

\bigskip

 \bigskip

 \bigskip

\newpage

\begin{figure}
{\bf(a)} \\
 \centerline{\epsfysize=0.6\linewidth \epsffile{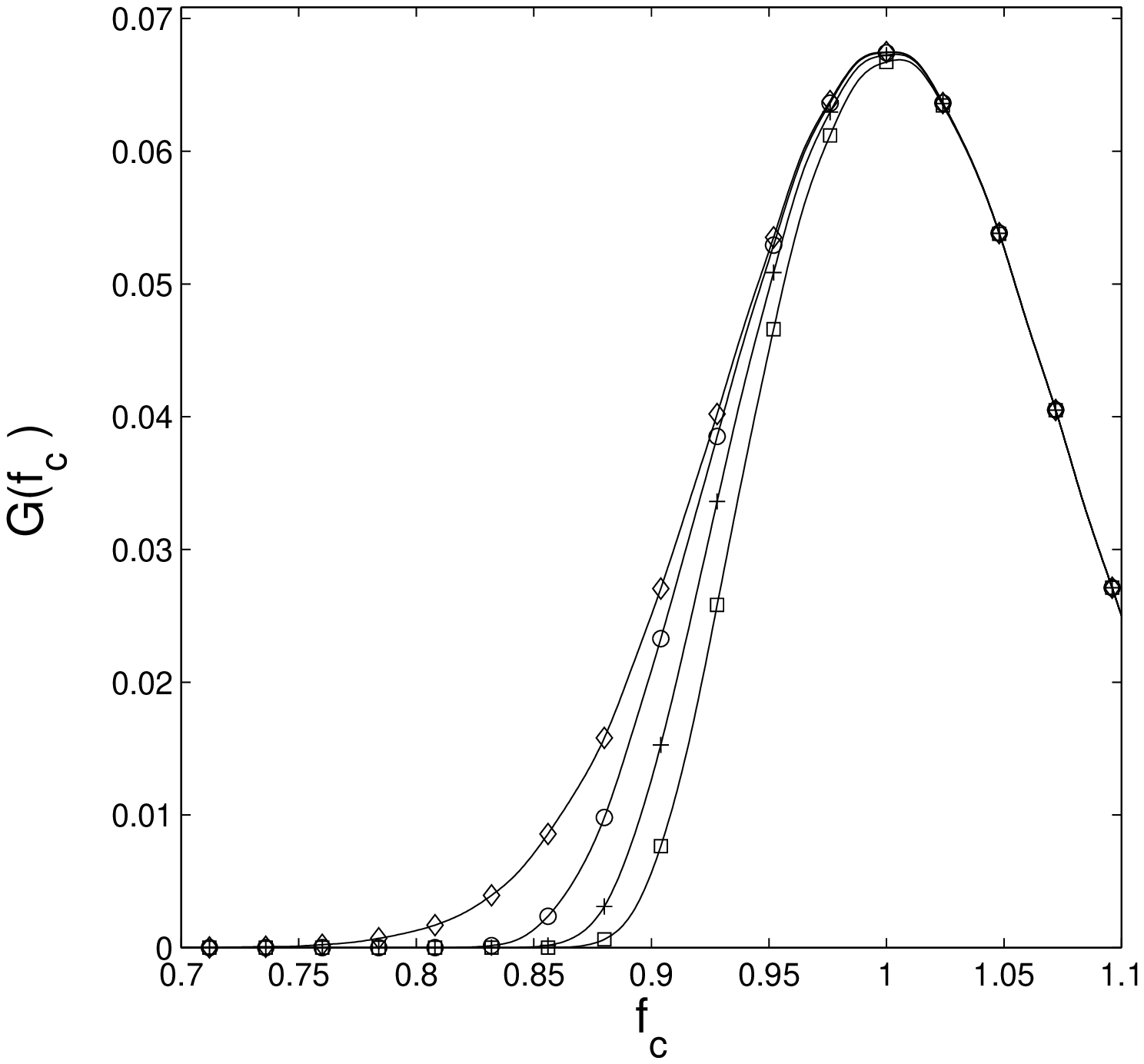}}
{\bf(b)} \\
 \centerline{\epsfysize=0.6\linewidth \epsffile{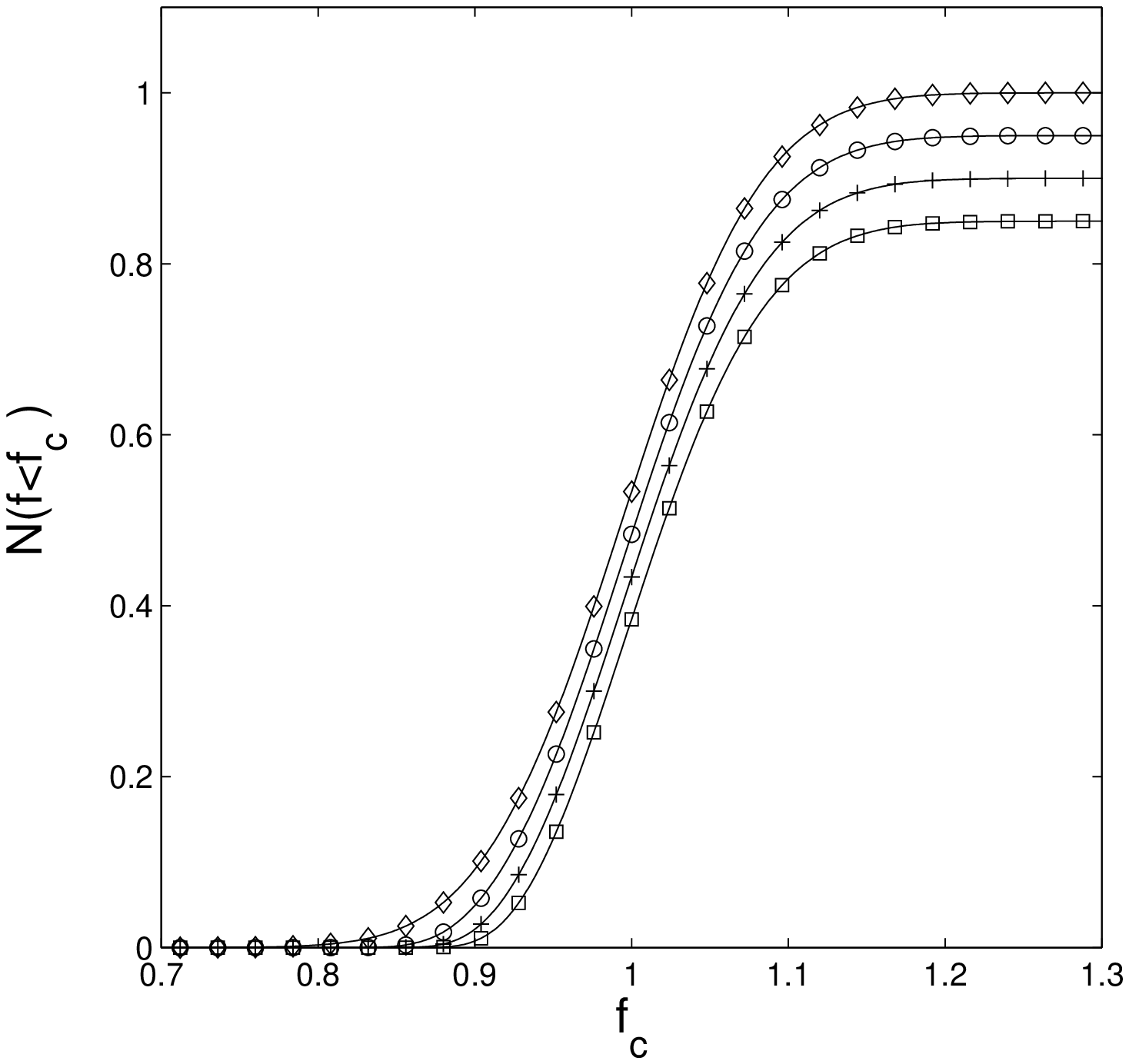}}
  \caption{Heterogeneous net.  Evolution as a function of $\phi$ of the
    threshold distribution at $f=0.6$,$KT=0.006 $ and $KTd=0.005$.
    (a) Evolution of $G(f_c) \ $ at
    $\phi=0 \ (\diamond), \phi=0.05 \ (\circ), \phi=0.1 \ (+)$ and
    $\phi=0.15 \ (\square)$. The fraction $\phi=0.15$ corresponds to $t\simeq 0.9 \tau$.
    (b) Evolution of the integral
    distribution $N(f<f_c)=\int_0^{f_c} G(x) dx$ at the same $\phi$ used in (a).
    }
    \label{verifica-Gauss}
\end{figure}


\begin{thebibliography}{99}
\bibitem{noise-comment}   A special comment has to be done on the source of the
  noise from a physical point of view. The simplest way of
  understanding this point is to remind that the DFBM has a one to
  one correspondence with a fuse network driven at imposed
  current. As a fuse is an electrical resistance the source of
  noise is the Nyquist noise on the resistance. In the case of the
  springs one can think that the springs have  a damping, which
   is a source of
  thermal noise because of the fluctuation dissipation theorem.
\end{thebibliography}
\end{document}